# Displacement and Squeeze Operators of a Three-Dimensional Harmonic Oscillator and Their Associated Quantum States

**Mehdi Miri, Sina Khorasani**

*School of Electrical Engineering*
*Sharif University of Technology*
*P. O. Box 11365-9363*
*Tehran, Iran*
Fax: +9821-6602-3261
Email: khorasani@sharif.edu

**Abstract**

We generalized the squeeze and displacement operators of the one-dimensional harmonic oscillator to the three-dimensional case and based on these operators we construct the corresponding coherent and squeezed states. We have also calculated the Wigner function for the three-dimensional harmonic oscillator and from the analysis of time evolution of this function, the quantum Liouville equation is also presented. Further properties of the quantum states including Mandel's $Q$ and quadrature squeezing parameters are discussed as well.

## 1. Introduction

The operator theory of harmonic oscillators [1] constitutes the groundwork of the elaborate quantum optical theory of photons. The quantization of electromagnetic radiation can be explained elegantly in terms of creator and annihilator operators, which operate on the corresponding energy levels [2-4]. Due to the second-order potential of harmonic oscillators, they can easily provide a direct bridge between classical optics and quantum optics through the phase-space Wigner functions [5], which are of extreme importance in the tomography of classical and non-classical lights [6-8].

Following the definition of coherent states put forward by Glauber [2-4], as the eigenstates of the annihilator operator, many studies have been done in order to generalize the concept of coherent states and the so-called squeezed states [5, 9]. Among these include an alternative definition of the generalized $k$-photon coherent states [10], which introduce a modification of the squeezing operator to describe higher-order interactions. In another report [11] the authors consider the generalization of coherent states and their superpositions connected through unitary transformations, where the transformation maps the ground state of the harmonic oscillator (vacuum state) onto an arbitrary superposition of $N \geq 2$ coherent states.

Since the successful demonstration of squeezed states of light in 1985 by Bell Laboratories [12], squeezed states have attracted much interest because of their possibility to significantly suppress the quantum noise, which is generally believed to be originated by the zero-point fluctuations of the vacuum [5]. Currently, squeezed states are routinely produced at laboratories using both solid-state and semiconductor lasers [13] and in high-$Q$ cavities [14].

Similarly, generalizations or extensions to the concept of squeezed states have been considered in numerous researches. Nieto [15] was the first to discuss the explicit functional forms for the squeeze and time-displacement operators and their applications, as successive multiplications of exponentials of simple operators. Bialynicki-Birula [16,17] presented a discussion of squeezed states of a generalized infinite-dimensional harmonic oscillator, when the ground state wave function takes on a Gaussian form. He furthermore presented the corresponding Wigner function and discussed its relativistic properties.

As another generalization of the simple one-dimensional harmonic oscillator, the problem of damped harmonic oscillator because of its time-dependent Hamiltonian was proposed and considered by Um et. al. [18], and they presented closed form expressions for squeeze and displacement operators. Also, Sohn and Swanson [19] have recently obtained exact transition elements of the squeezed harmonic oscillator when the generalized Hamiltonian describes two-photon processes, using Bogoliubov transformations. Fakhri [20] considered the three-dimensional (3D) harmonic oscillator and Morse potentials, and showed that the constructed Heisenberg Lie superalgebras would lead to multiple supercharges. In his analysis, he analyzed the 3D harmonic oscillator in the spherical system of coordinates. Finally, Fan and Jiang [21] have constructed three mutually commuting squeeze operators, which are applicable to three-mode states.

In this paper, we revisit the 3D harmonic oscillator and obtain generalized expressions for the corresponding coherent and squeezed states, starting from the Cartesian coordinates in which the harmonic oscillator can be easily factorized. We also present closed-form simple expressions which explicitly represent the corresponding displacement and squeeze operators, and the corresponding generalized Mandel's $Q$ parameter is obtained for the generalized squeezed state in the form of a vector. We show that how proper definition of vector operators and variable could greatly simplify the notations of operators and eigenstates.

## 2. Coherent states and the displacement operator

### 2.1. Wigner function for 3D harmonic oscillator

We can calculate wave function of three-dimensional (3D) harmonic oscillator directly from the Schrödinger equation, with the diagonalized potential given by

$$U(\mathbf{r}) = \frac{M}{2}\left(\omega_{xx}^{2}x^{2} + \omega_{yy}^{2}y^{2} + \omega_{zz}^{2}z^{2}\right) \quad (2.1)$$

Here, without loss of generality one may assume that $\omega_{xx} = \omega_{yy} = \omega_{zz} = \omega$. Now let $|m,n,l\rangle$ denote the energy eigenstates of 3D harmonic oscillator, hence for the corresponding annihilation and creation operators we have

$$\hat{a}_x |m,n,l\rangle = \sqrt{m}|m-1,n,l\rangle \quad (2.2-\text{a})$$
$$\hat{a}_y |m,n,l\rangle = \sqrt{n}|m,n-1,l\rangle \quad (2.2-\text{b})$$
$$\hat{a}_z |m,n,l\rangle = \sqrt{l}|m,n,l-1\rangle \quad (2.2-\text{c})$$
$$\hat{a}_x^\dagger |m,n,l\rangle = \sqrt{m+1}|m+1,n,l\rangle \quad (2.2-\text{d})$$
$$\hat{a}_y^\dagger |m,n,l\rangle = \sqrt{n+1}|m,n+1,l\rangle \quad (2.2-\text{e})$$
$$\hat{a}_z^\dagger |m,n,l\rangle = \sqrt{l+1}|m,n,l+1\rangle \quad (2.2-\text{f})$$

Hence, the eigenfunctions will be

$$\Psi_{nml}(\mathbf{r}) = \langle \mathbf{r}|m,n,l\rangle = \frac{1}{\sqrt{2^{n+m+l}n!\,m!\,l!}} \left(\frac{\kappa^2}{\pi}\right)^{\frac{3}{4}} \exp\left(-\frac{1}{2}\kappa^2 r^2\right) H_n(\kappa x) H_m(\kappa y) H_l(\kappa z) \qquad (2.3)$$

where $\kappa = \sqrt{M\omega/\hbar}$. We can find the corresponding Wigner function for this system from the definition of the Wigner function as

$$W_{|n,m,l\rangle}(\mathbf{r},\mathbf{p}) = \left(\frac{1}{2\pi\hbar}\right)^3 \iiint_{-\infty}^{\infty} d^3\zeta \exp\left(-\frac{i}{\hbar}\mathbf{p}\cdot\boldsymbol{\zeta}\right) \langle \mathbf{r}+\tfrac{1}{2}\boldsymbol{\zeta}|\hat{\rho}|\mathbf{r}-\tfrac{1}{2}\boldsymbol{\zeta}\rangle \qquad (2.4)$$

Here, $\hat{\rho}$ is the density operator and $\boldsymbol{\zeta} = \zeta_x \mathbf{i} + \zeta_y \mathbf{j} + \zeta_z \mathbf{k}$ represents the dummy integration variable. In the case of pure state with $\hat{\rho} = |m,n,l\rangle\langle m,n,l|$ gives

$$W_{|n,m,l\rangle}(\mathbf{r},\mathbf{p}) = \left(\frac{1}{2\pi\hbar}\right)^3 \iiint_{-\infty}^{\infty} d^3\zeta \exp\left(-\frac{i}{\hbar}\mathbf{p}\cdot\boldsymbol{\zeta}\right) \Psi_{nml}^*\left(\mathbf{r}-\tfrac{1}{2}\boldsymbol{\zeta}\right) \Psi_{nml}\left(\mathbf{r}+\tfrac{1}{2}\boldsymbol{\zeta}\right) \qquad (2.5)$$

The Wigner function of 3D harmonic oscillator will take the form

$$W_{|n,m,l\rangle}(\mathbf{r},\mathbf{p}) = \frac{(-1)^{n+m+l}}{(\pi\hbar)^3} \exp\left[-\left(\frac{\mathbf{p}}{\hbar\kappa}\right)^2 - (\kappa\mathbf{r})^2\right]$$
$$L_n\left\{2\left[\left(\frac{p_x}{\hbar\kappa}\right)^2 + (\kappa x)^2\right]\right\} L_m\left\{2\left[\left(\frac{p_y}{\hbar\kappa}\right)^2 + (\kappa y)^2\right]\right\} L_l\left\{2\left[\left(\frac{p_z}{\hbar\kappa}\right)^2 + (\kappa z)^2\right]\right\} \qquad (2.6)$$

in which $L_n(x)$ is the Laguerre function of order $n$; see appendix A for the detailed derivation of (2.6). For the generation of coherent states, we must apply a suitable displacement operator to the ground state of 3D harmonic oscillator. In doing so, we need to generalize the method of [15] in construction of 3D displacement operator.

*2.2. Construction of coherent state*

The ground state of a 3D harmonic oscillator is given by

$$\Psi_{\mathbf{0}}(\mathbf{r}) = \langle \mathbf{r}|\mathbf{0}\rangle = \left(\frac{\kappa^2}{\pi}\right)^{\frac{3}{4}} \exp\left(-\frac{1}{2}\kappa^2 r^2\right) \quad (2.7)$$

in which the ground state $|\mathbf{0}\rangle$ is defined using the null integer triplet $\mathbf{0} = (0,0,0)$. Now we define the displacement operator as

$$\widehat{D}(\boldsymbol{\alpha}) = \exp(\alpha_x \hat{a}_x{}^\dagger - \alpha_x{}^* \hat{a}_x)\exp(\alpha_y \hat{a}_y{}^\dagger - \alpha_y{}^* \hat{a}_y)\exp(\alpha_z \hat{a}_z{}^\dagger - \alpha_z{}^* \hat{a}_z) \quad (2.8)$$

Here, the displacement vector $\boldsymbol{\alpha} = \alpha_x \mathbf{i} + \alpha_y \mathbf{j} + \alpha_z \mathbf{k}$, with $\alpha_x$, $\alpha_y$, and $\alpha_z$ being complex constants. As will be shown, the order of displacements along $x$, $y$, and $z$ is irrelevant. This is because of the obvious relations

$$[\hat{a}_\iota, \hat{a}_\nu] = [\hat{a}_\iota{}^\dagger, \hat{a}_\nu{}^\dagger] = 0, \quad \iota, \nu = x, y, z \quad (2.9-a)$$

$$[\hat{a}_\iota, \hat{a}_\nu{}^\dagger] = [\hat{a}_\iota{}^\dagger, \hat{a}_\nu] = 0, \quad \iota \neq \nu \quad (2.9-b)$$

In trying to find a compact form for this operator we start from the Baker-Campbell-Hausdorff relation [5], which reads

$$\exp(\hat{A} + \hat{B}) = \exp(\hat{A})\exp(\hat{B})\exp\left(\frac{1}{2}[\hat{A}, \hat{B}]\right) \quad (2.10)$$

given that

$$\left[\hat{A}, [\hat{A}, \hat{B}]\right] = \left[\hat{B}, [\hat{A}, \hat{B}]\right] = 0 \quad (2.11)$$

Hence the displacement operator is simplified into the compact form

$$\widehat{D}(\boldsymbol{\alpha}) = \exp(\boldsymbol{\alpha} \cdot \hat{\mathbf{a}}^\dagger - \boldsymbol{\alpha}^* \cdot \hat{\mathbf{a}}) \quad (2.12)$$

where the vector creation and annihilation operators are defined by

$$\hat{\mathbf{a}} = \hat{a}_x \mathbf{i} + \hat{a}_y \mathbf{j} + \hat{a}_z \mathbf{k} \quad (2.13-a)$$

$$\hat{\mathbf{a}}^\dagger = \hat{a}_x{}^\dagger \mathbf{i} + \hat{a}_y{}^\dagger \mathbf{j} + \hat{a}_z{}^\dagger \mathbf{k} \quad (2.13-b)$$

The application of the displacement operator $\hat{D}(\boldsymbol{\alpha})$ to the ground state $|0\rangle$ results in

$$\hat{D}(\boldsymbol{\alpha})|0\rangle = |\boldsymbol{\alpha}\rangle \quad (2.14)$$

where $|\boldsymbol{\alpha}\rangle$ is defined as the generalized coherent state in 3D. Also, from the properties of $\hat{\mathbf{a}}$ and $\hat{\mathbf{a}}^\dagger$ one can further observe that

$$\hat{D}(\boldsymbol{\alpha}) = \exp\left(-\frac{1}{2}\boldsymbol{\alpha}\cdot\boldsymbol{\alpha}^*\right)\exp(\boldsymbol{\alpha}\cdot\hat{\mathbf{a}}^\dagger)\exp(-\boldsymbol{\alpha}^*\cdot\hat{\mathbf{a}}) \quad (2.15)$$

The position representation of $|\boldsymbol{\alpha}\rangle$ will be

$$\langle\mathbf{r}|\boldsymbol{\alpha}\rangle = \left(\frac{\kappa^2}{\pi}\right)^{\frac{3}{4}} \exp\left\{-\frac{1}{2}\left[(\kappa x - \sqrt{2}\alpha_x)^2 + (\kappa y - \sqrt{2}\alpha_y)^2 + (\kappa z - \sqrt{2}\alpha_z)^2\right]\right\}$$

$$= \left(\frac{\kappa^2}{\pi}\right)^{\frac{3}{4}} \exp\left(-\frac{1}{2}|\kappa\mathbf{r} - \sqrt{2}\boldsymbol{\alpha}|^2\right) \quad (2.16)$$

The direct application of the displacement operator also can be simply shown to equally result in the triple infinite series of the generalized coherent state as

$$|\boldsymbol{\alpha}\rangle = \hat{D}(\boldsymbol{\alpha})|0\rangle = \exp\left(-\frac{1}{2}\boldsymbol{\alpha}^*\cdot\boldsymbol{\alpha}\right) \sum_{m,n,l=0}^{\infty} \frac{1}{m!\,n!\,l!} (\alpha_x\mathbf{i}\cdot\hat{\mathbf{a}}^\dagger)^m (\alpha_y\mathbf{j}\cdot\hat{\mathbf{a}}^\dagger)^n (\alpha_z\mathbf{k}\cdot\hat{\mathbf{a}}^\dagger)^l |0\rangle$$

$$= \exp\left(-\frac{1}{2}\boldsymbol{\alpha}^*\cdot\boldsymbol{\alpha}\right) \sum_{m,n,l=0}^{\infty} \frac{\alpha_x^m \alpha_y^n \alpha_z^l}{m!\,n!\,l!} |m,n,l\rangle \quad (2.17)$$

*2.3. Over-completeness of coherent states*

As one of the important properties of coherent sates we can examine the over-completeness of the proposed coherent sates. A set of states are called over-complete if they form a complete set and are not orthogonal. We first consider the completeness of coherent states:

$$\int_{-\infty}^{\infty} |\boldsymbol{\alpha}\rangle\langle\boldsymbol{\alpha}|\, d^6\alpha = \iiint_{-\infty}^{\infty} |\boldsymbol{\alpha}\rangle\langle\boldsymbol{\alpha}|\, d^2\alpha_x\, d^2\alpha_y\, d^2\alpha_z = \iiint_{-\infty}^{\infty} \exp\left(-\frac{1}{2}\boldsymbol{\alpha}^*\cdot\boldsymbol{\alpha}\right) \exp\left(-\frac{1}{2}\boldsymbol{\alpha}\cdot\boldsymbol{\alpha}^*\right)$$

$$\sum_{n,m,l=0}^{\infty} \sum_{p,q,w=0}^{\infty} \frac{\alpha_x^n \alpha_x^{*p} \alpha_y^m \alpha_y^{*q} \alpha_z^l \alpha_z^{*w} |n,m,l\rangle\langle p,q,w|}{\sqrt{n!\,m!\,l!\,p!\,q!\,w!}} d^2\alpha_x\, d^2\alpha_y\, d^2\alpha_z \qquad (2.18)$$

where $d^2\alpha_\iota = d(\text{Re}\{\alpha_\iota\})\, d(\text{Im}\{\alpha_\iota\})$ ; $\iota = x, y, z$. Using the change of variables:

$$\begin{cases} \alpha_\iota = r_\iota \exp(i\theta_\iota) \\ d^2\alpha_\iota = r_\iota dr_\iota d\theta_\iota \end{cases} ; \quad \iota = x, y, z \qquad (2.19)$$

results in:

$$\sum_{n,m,l=0}^{\infty} \sum_{p,q,w=0}^{\infty} \frac{|n,m,l\rangle\langle p,q,w|}{\sqrt{n!\,m!\,l!\,p!\,q!\,w!}} \iiint_{-\infty}^{\infty} r_x^{n+p+1} r_y^{m+q+1} r_z^{l+w+1} \exp(-r^2)\, dr_x\, dr_y\, dr_z$$

$$\int_0^{2\pi}\int_0^{2\pi}\int_0^{2\pi} \exp\{i[(n-p)\theta_x + (m-q)\theta_y + (l-w)\theta_z]\}\, d\theta_x\, d\theta_y\, d\theta_z \quad ; \quad r^2 = r_x^2 + r_y^2 + r_z^2 \qquad (2.20)$$

It is known that $\int_0^{2\pi} \exp[i(n-m)\theta]\, d\theta = 2\pi\delta_{nm}$, so using $r_\iota^2 = \gamma_\iota \to 2r_\iota dr_\iota = d\gamma_\iota$ ; $\iota = x, y, z$, we get:

$$\int_{-\infty}^{\infty} |\boldsymbol{\alpha}\rangle\langle\boldsymbol{\alpha}|\, d^6\alpha =$$

$$\pi^3 \sum_{n,m,l=0}^{\infty} \frac{|n,m,l\rangle\langle n,m,l|}{n!\,m!\,l!} \int_{-\infty}^{\infty} \exp(-\gamma_x)\gamma_x^n\, d\gamma_x \int_{-\infty}^{\infty} \exp(-\gamma_y)\gamma_y^m\, d\gamma_y \int_{-\infty}^{\infty} \exp(-\gamma_z)\gamma_z^l\, d\gamma_z \qquad (2.21)$$

which can be easily simplified by using the identity $\int_{-\infty}^{\infty} \exp(-\gamma)\gamma^n\, d\gamma = n!$ into the expression:

$$\int_{-\infty}^{\infty} |\boldsymbol{\alpha}\rangle\langle\boldsymbol{\alpha}|\, d^6\alpha = \pi^3 \sum_{n,m,l=0}^{\infty} |n,m,l\rangle\langle n,m,l| = \pi^3 \qquad (2.22)$$

Therefore the proposed coherent sates constitute a complete set. Now we examine their non-orthogonality by considering the inner product of two different coherent states $|\boldsymbol{\alpha}\rangle$ and $|\boldsymbol{\beta}\rangle$ :

$$\langle\boldsymbol{\beta}|\boldsymbol{\alpha}\rangle = \exp\left(-\frac{1}{2}\boldsymbol{\alpha}^*\cdot\boldsymbol{\alpha}\right)\exp\left(-\frac{1}{2}\boldsymbol{\beta}^*\cdot\boldsymbol{\beta}\right)\sum_{n,m,l=0}^{\infty}\sum_{p,q,w=0}^{\infty}\frac{\beta_x^{*n}\beta_y^{*m}\beta_z^{*l}\,\alpha_x^p\alpha_y^q\alpha_z^w\,\langle n,m,l|p,q,w\rangle}{\sqrt{n!\,m!\,l!\,p!\,q!\,w!}}$$

$$= \exp\left[-\frac{1}{2}(\boldsymbol{\alpha}^*\cdot\boldsymbol{\alpha}+\boldsymbol{\beta}^*\cdot\boldsymbol{\beta})\right]\sum_{n,m,l=0}^{\infty}\frac{(\beta_x^*\alpha_x)^n(\beta_y^*\alpha_y)^m(\beta_z^*\alpha_z)^l}{n!\,m!\,l!}$$

$$= \exp\left[-\frac{1}{2}(\boldsymbol{\alpha}^*\cdot\boldsymbol{\alpha}+\boldsymbol{\beta}^*\cdot\boldsymbol{\beta})+\boldsymbol{\beta}^*\cdot\boldsymbol{\alpha}\right]$$

$$= \exp\left[\frac{1}{2}(\boldsymbol{\beta}^*\cdot\boldsymbol{\alpha}-\boldsymbol{\beta}\cdot\boldsymbol{\alpha}^*)\right]\exp\left[-\frac{1}{2}(\boldsymbol{\beta}^*-\boldsymbol{\alpha}^*)\cdot(\boldsymbol{\beta}-\boldsymbol{\alpha})\right] \qquad (2.23)$$

Hence, we obtain the squared magnitude of the inner product as:

$$|\langle\boldsymbol{\beta}|\boldsymbol{\alpha}\rangle|^2 = \exp[-(\boldsymbol{\beta}-\boldsymbol{\alpha})^*\cdot(\boldsymbol{\beta}-\boldsymbol{\alpha})] \neq 0 \qquad (2.24)$$

which declares that 3D coherent states are not orthogonal, but their inner product tends to vanish, when $|\boldsymbol{\beta}-\boldsymbol{\alpha}|$ is sufficiently large. Equations (2.22) and (2.24), establish, therefore, that the proposed coherent sates are over-complete.

*2.4. Quantum Liouville equation in 3D*

Now we attempt to find the quantum Liouville equation for the generalized 3D harmonic oscillator system of interest. So we start from time evolution of our six-dimensional (6D) Wigner function. We have the Von Neumann equation for the time evolution of density operator as [1,5]

$$\frac{\partial\hat{\rho}}{\partial t} = -\frac{i}{\hbar}[\hat{H},\hat{\rho}] \qquad (2.25)$$

with $\hat{H}$ representing the Hamiltonian of the 3D harmonic oscillator. Starting from this equation we can show that

$$\frac{\partial}{\partial t}\left\langle\mathbf{r}+\frac{1}{2}\boldsymbol{\zeta}\bigg|\hat{\rho}\bigg|\mathbf{r}-\frac{1}{2}\boldsymbol{\zeta}\right\rangle = -\frac{i}{\hbar}\left\langle\mathbf{r}+\frac{1}{2}\boldsymbol{\zeta}\bigg|[\hat{H},\hat{\rho}]\bigg|\mathbf{r}-\frac{1}{2}\boldsymbol{\zeta}\right\rangle \qquad (2.26)$$

With substitution of (2.19) in the definition of Wigner function we get

$$\frac{\partial}{\partial t}W(\mathbf{r},\mathbf{p},t) = T + Q \qquad (2.27)$$

Here, $T$ and $Q$ respectively correspond to kinetic and potential energies in terms of Moyal functions and introduce a Fourier transform as in [5]. The derivation closely follows the approach in [5], however, we employ the generalized 3D expressions for functions and operators. Hence, $T$ and $Q$ will be given by the expressions

$$T = -\frac{i}{\hbar}\frac{1}{2M}\left(\frac{1}{2\pi\hbar}\right)^3 \iiint_{-\infty}^{\infty} d^3\zeta \exp\left(-\frac{i}{\hbar}\mathbf{p}\cdot\boldsymbol{\zeta}\right) G(\mathbf{r},\mathbf{p},\boldsymbol{\zeta})$$

$$G(\mathbf{r},\mathbf{p},\boldsymbol{\zeta}) = \left\langle \mathbf{r}+\frac{1}{2}\boldsymbol{\zeta}\middle|[\hat{\mathbf{p}}^2,\hat{\rho}]\middle|\mathbf{r}-\frac{1}{2}\boldsymbol{\zeta}\right\rangle \qquad (2.28-\text{a})$$

and

$$Q = -\frac{i}{\hbar}\left(\frac{1}{2\pi\hbar}\right)^3 \iiint_{-\infty}^{\infty} d^3\zeta \exp\left(-\frac{i}{\hbar}\mathbf{p}\cdot\boldsymbol{\zeta}\right) R(\mathbf{r},\mathbf{p},\boldsymbol{\zeta})$$

$$R(\mathbf{r},\mathbf{p},\boldsymbol{\zeta}) = \left\langle \mathbf{r}+\frac{1}{2}\boldsymbol{\zeta}\middle|[\hat{U},\hat{\rho}]\middle|\mathbf{r}-\frac{1}{2}\boldsymbol{\zeta}\right\rangle \qquad (2.28-\text{b})$$

The potential energy operator $\hat{U}$ is similarly defined as in [5]. Now we need to calculate $T$ and $Q$ in terms of the Wigner function and its higher-order derivatives. First we consider the kinetic energy term, which gives rise after some mathematical manipulations to the following equation for the kinetic energy term

$$T = -\frac{1}{M}\mathbf{p}\cdot\nabla_r W(\mathbf{r},\mathbf{p},t) \qquad (2.29)$$

Now consider potential energy term. From the 3D Taylor expansion near $\mathbf{r}$, we have:

$$U\left(\mathbf{r}\pm\frac{1}{2}\boldsymbol{\zeta}\right) = \sum_{n=0}^{\infty}\frac{1}{n!}\left(\pm\frac{1}{2}\boldsymbol{\zeta}\cdot\nabla\right)^n U(\mathbf{r}) \qquad (2.30)$$

Therefore we obtain

$$U\left(\mathbf{r}+\frac{1}{2}\boldsymbol{\zeta}\right)+U\left(\mathbf{r}-\frac{1}{2}\boldsymbol{\zeta}\right) = \sum_{n=0}^{\infty}\frac{(i\hbar)^{2n}}{2^{2n}(2n)!}\left(-\frac{i}{\hbar}\right)^{2n}(\boldsymbol{\zeta}\cdot\nabla)^{2n}U(\mathbf{r}) \qquad (2.31)$$

and

$$U\left(\mathbf{r}+\frac{1}{2}\boldsymbol{\zeta}\right)-U\left(\mathbf{r}-\frac{1}{2}\boldsymbol{\zeta}\right) = 2\sum_{n=0}^{\infty}\frac{(i\hbar)^{2n+1}}{2^{2n+1}(2n+1)!}\left(-\frac{i}{\hbar}\right)^{2n+1}(\boldsymbol{\zeta}\cdot\nabla)^{2n+1}U(\mathbf{r}) \qquad (2.32)$$

We finally have the closed-form expression

$$Q = \sum_{n=0}^{\infty}\frac{(-1)^n\hbar^{2n}}{2^{2n}(2n+1)!}(\nabla\cdot\nabla_p)^{2n+1}U(\mathbf{r})W(\mathbf{r},\mathbf{p},t) \qquad (2.33)$$

With substitution of (2.29) and (2.33) in (2.27) we get the Liouville's equation for the time-evolution of the Wigner function in 3D in the compact form

$$\left[\frac{\partial}{\partial t}+\frac{1}{M}\mathbf{p}\cdot\nabla\right]W(\mathbf{r},\mathbf{p},t) = \sum_{n=0}^{\infty}\frac{(-1)^n\hbar^{2n}}{2^{2n}(2n+1)!}(\nabla\cdot\nabla_p)^{2n+1}U(\mathbf{r})W(\mathbf{r},\mathbf{p},t) \qquad (2.34)$$

For the 3D harmonic oscillator in general case, the potential $U(\mathbf{r})$ is second-order in $\mathbf{r}$. So for the harmonic oscillator, the right-hand-side of the Liouville's equation is equal to zero. Hence, the quantum Liouville's equation for 3D harmonic oscillator will be simply

$$\left[\frac{\partial}{\partial t}+\frac{1}{M}\mathbf{p}\cdot\nabla-\nabla U(\mathbf{r})\cdot\nabla_p\right]W(\mathbf{r},\mathbf{p},t) = 0 \qquad (2.35)$$

*2.5. Time Evolution of Coherent State*

The Hamiltonian of 3D harmonic oscillator is time independent, so for the temporal evolution of the coherent state we can write down

$$|\Psi_{\text{coh}}(t)\rangle = \exp\left(-\frac{i}{\hbar}\widehat{H}t\right)\sum_{m,n,l=0}^{\infty}|m,n,l\rangle\langle m,n,l|\Psi_{\text{coh}}(0)\rangle \qquad (2.36)$$

Defining the expansion coefficients as

$$\Omega_{m,n,l} = \langle m, n, l | \Psi_{\text{coh}}(0) \rangle \quad (2.37)$$

we get after some algebra

$$\Omega_{m,n,l} = \exp\left(-\frac{1}{2}\boldsymbol{\alpha}^* \cdot \boldsymbol{\alpha}\right) \frac{\alpha_x^m \alpha_y^n \alpha_z^l}{\sqrt{m!\,n!\,l!}} \quad (2.38)$$

On the other hand

$$\hat{H} |m, n, l\rangle = \hbar\omega \left(m + n + l + \frac{3}{2}\right) |m, n, l\rangle \quad (2.39)$$

After simplifying we have

$$|\Psi_{\text{coh}}(t)\rangle = \exp\left(-\frac{3}{2}i\omega t\right) \sum_{m,n,l=0}^{\infty} \frac{(\alpha_x e^{-i\omega t})^m (\alpha_y e^{-i\omega t})^n (\alpha_z e^{-i\omega t})^l}{\sqrt{m!\,n!\,l!}}$$

$$\exp\left[-\frac{1}{2}(e^{-i\omega t}\boldsymbol{\alpha})^* \cdot (e^{-i\omega t}\boldsymbol{\alpha})\right] |m, n, l\rangle \quad (2.40)$$

from which we obtain

$$|\Psi_{\text{coh}}(t)\rangle = \exp\left(-\frac{3}{2}i\omega t\right) |\Psi_{\text{coh}} e^{-i\omega t}\rangle \quad (2.41)$$

From this relation we see that the time evolution of coherent state is also a coherent state, and also after one period $\frac{2\pi}{\omega}$ of oscillation, the state vector phase change is $\frac{3}{2}\omega \times \frac{2\pi}{\omega} = 3\pi$.

*2.6. Position representation of coherent state*

In the below we try to find a compact form for position representation of coherent state of 3D harmonic oscillator. This results in

$\langle \mathbf{r} | \Psi_{\text{coh}}(t) \rangle =$

$$\left(\frac{\kappa^2}{\pi}\right)^{\frac{3}{4}} \exp\left[-\frac{1}{2}(3i\omega t + \boldsymbol{\alpha}^* \cdot \boldsymbol{\alpha} + \kappa^2 r^2)\right] \exp\left(-\frac{1}{2} e^{-2i\omega t} \boldsymbol{\alpha} \cdot \boldsymbol{\alpha}\right) \exp(\sqrt{2} e^{-i\omega t} \kappa \mathbf{r} \cdot \boldsymbol{\alpha}) \qquad (2.42)$$

Using the definitions

$$\bar{\mathbf{r}}(t) = \frac{\sqrt{2}}{\kappa} \text{Re}\{e^{-i\omega t} \boldsymbol{\alpha}\} \qquad (2.43 - a)$$

$$\bar{\mathbf{p}}(t) = \sqrt{2} \hbar \kappa \text{Im}\{e^{-i\omega t} \boldsymbol{\alpha}\} \qquad (2.43 - b)$$

$$\Phi_{\text{zp}}(t) = \frac{3}{2}\omega t \qquad (2.43 - c)$$

$$iA_\Delta(t) = \frac{\boldsymbol{\alpha}^* \cdot \boldsymbol{\alpha}}{2}(1 + e^{-2i\omega t}) - \frac{\kappa^2}{2} \bar{\mathbf{r}}(t) \cdot \bar{\mathbf{r}}(t) = \frac{1}{2}\left(e^{-2i\omega t} \boldsymbol{\alpha}^* \cdot \boldsymbol{\alpha} - \text{Re}\{e^{-2i\omega t} \boldsymbol{\alpha} \cdot \boldsymbol{\alpha}\}\right) \qquad (2.43 - d)$$

Finally, the complete form of position space representation of coherent state is

$$\Psi_{\text{coh}}(\mathbf{r}, t) = \langle \mathbf{r} | \Psi_{\text{coh}}(t) \rangle$$

$$= \left(\frac{\kappa^2}{\pi}\right)^{\frac{3}{4}} \exp\left\{-\frac{\kappa^2}{2}[\mathbf{r} - \bar{\mathbf{r}}(t)] \cdot [\mathbf{r} - \bar{\mathbf{r}}(t)]\right\} \exp\left[\frac{i}{\hbar} \bar{\mathbf{p}}(t) \cdot \bar{\mathbf{r}}(t)\right] \exp\{-i[\Phi_{\text{zp}}(t) + A_\Delta(t)]\} \qquad (2.44)$$

Here, as for the case in 1D problem $A_\Delta(t)$ for real $\boldsymbol{\alpha}$ will take on real values, otherwise it will be complex.

### 3. Squeezed states and the squeeze operator

Now we try to find functional form of squeeze operator and squeezed state of 3D harmonic oscillator and position representation of this squeezed state. For 1D case, the squeeze operator is defined as

$$\hat{S}(s) = \exp\left(\frac{1}{2} s \hat{a}^\dagger \hat{a}^\dagger - \frac{1}{2} s^* \hat{a} \hat{a}\right) \qquad (3.1)$$

where in general $s$ is a complex number

$$s = |s|e^{i\theta} = s_1 + is_2 \qquad (3.2)$$

Expanded form of 1D squeezing operator is therefore

$$\hat{S}(s) = \exp\left[\frac{1}{2}e^{i\theta}\tanh|s|(\hat{a}^\dagger)^2\right]\operatorname{sech}^{\frac{1}{2}}|s|\left[\sum_{n=0}^{\infty}\frac{(\operatorname{sech}|s|-1)^n}{n!}(\hat{a}^\dagger)^n(\hat{a})^n\right]\exp\left[-\frac{1}{2}e^{-i\theta}\tanh|s|(\hat{a})^2\right] \qquad (3.3)$$

If we use notation of [15] as

$$\hat{x} = \frac{1}{\sqrt{2}}(\hat{a} + \hat{a}^\dagger) \qquad (3.4-a)$$

$$\hat{\partial} = i\hat{p} = \frac{1}{\sqrt{2}}(\hat{a} - \hat{a}^\dagger) \qquad (3.4-b)$$

the new form of squeezing operator will take the form

$$\hat{S}(s) = \exp\left[-s_1\left(\hat{x}\hat{\partial} + \frac{1}{2}\right) + \frac{i}{2}s_2(\hat{x}^2 + \hat{\partial}^2)\right] \qquad (3.5)$$

In the expanded form we have

$$\hat{S}(s) = g^{-\frac{1}{2}}\exp\left[i\frac{s_2}{2|s|}\frac{\sinh|s|}{g}\hat{x}^2\right]\exp[-\ln(g)\hat{x}\hat{\partial}]\exp\left[i\frac{s_2}{2|s|}\frac{\sinh|s|}{g}\hat{\partial}^2\right] \qquad (3.6)$$

where

$$g = \cosh|s| + \frac{s_1}{|s|}\sinh|s| = e^{|s|}\cos^2\left(\frac{\theta}{2}\right) + e^{-|s|}\sin^2\left(\frac{\theta}{2}\right) \qquad (3.7)$$

In generalization of this concept to 3D case we consider 3D squeezing as independently squeezing of wave function of harmonic oscillator in the three $x$, $y$, and $z$ dimensions. So for the 3D harmonic oscillator this method can be applied directly resulting as

$$\hat{S}(\mathbf{s}) = \hat{S}_x(s_x)\hat{S}_y(s_y)\hat{S}_z(s_z) \qquad (3.8)$$

Here, $\hat{S}_x(s_x)$, $\hat{S}_y(s_y)$, and $\hat{S}_z(s_z)$ operate only on $x$, $y$, and $z$ dimensions, respectively, and hence the order of their appearance is irrelevant as will be shown shortly. We have

$$\hat{S}_x(s_x) = \exp\left(\frac{1}{2}s_x \hat{a}_x^\dagger \hat{a}_x^\dagger - \frac{1}{2}s_x^* \hat{a}_x \hat{a}_x\right) \qquad (3.9-a)$$

$$\hat{S}_y(s_y) = \exp\left(\frac{1}{2}s_y \hat{a}_y^\dagger \hat{a}_y^\dagger - \frac{1}{2}s_y^* \hat{a}_y \hat{a}_y\right) \qquad (3.9-b)$$

$$\hat{S}_z(s_z) = \exp\left(\frac{1}{2}s_z \hat{a}_z^\dagger \hat{a}_z^\dagger - \frac{1}{2}s_z^* \hat{a}_z \hat{a}_z\right) \qquad (3.9-c)$$

Therefore

$$\hat{S}(\mathbf{s}) =$$

$$\exp\left[\frac{1}{2}s_x(\hat{a}_x^\dagger)^2 - \frac{1}{2}s_x^*(\hat{a}_x)^2\right] \exp\left[\frac{1}{2}s_y(\hat{a}_y^\dagger)^2 - \frac{1}{2}s_y^*(\hat{a}_y)^2\right] \exp\left[\frac{1}{2}s_z(\hat{a}_z^\dagger)^2 - \frac{1}{2}s_z^*(\hat{a}_z)^2\right] \qquad (3.10)$$

Now let the following definitions hold

$$\hat{v} = \frac{1}{\sqrt{2}}(\hat{a}_v + \hat{a}_v^\dagger) \qquad (3.11-a)$$

$$\hat{\partial}_v = i\hat{p}_v = \frac{1}{\sqrt{2}}(\hat{a}_v - \hat{a}_v^\dagger) \qquad (3.11-b)$$

in which $v = x, y, z$. We furthermore we can show that

$$\left[(\hat{a}_\iota)^2, (\hat{a}_v^\dagger)^2\right] = \left[(\hat{a}_\iota^\dagger)^2, (\hat{a}_v)^2\right] = [(\hat{a}_\iota)^2, (\hat{a}_v)^2] = \left[(\hat{a}_\iota^\dagger)^2, (\hat{a}_v^\dagger)^2\right] = 0 \qquad (3.12)$$

where

$$\iota, v = x, y, z$$
$$\iota \neq v$$

With subsequent use of (3.12) we can show that

$$\left[(\hat{a}_\iota^\dagger)^2, \left[(\hat{a}_\iota^\dagger)^2, (\hat{a}_v)^2\right]\right] = \left[(\hat{a}_v)^2, \left[(\hat{a}_\iota^\dagger)^2, (\hat{a}_v)^2\right]\right] = 0 \qquad (3.13-a)$$

$$\left[(\hat{a}_\iota)^2, \left[(\hat{a}_\iota)^2, (\hat{a}_\nu{}^\dagger)^2\right]\right] = \left[(\hat{a}_\nu{}^\dagger)^2, \left[(\hat{a}_\iota)^2, (\hat{a}_\nu{}^\dagger)^2\right]\right] = 0 \qquad (3.13-b)$$

$$[(\hat{a}_\iota)^2, [(\hat{a}_\iota)^2, (\hat{a}_\nu)^2]] = [(\hat{a}_\nu)^2, [(\hat{a}_\iota)^2, (\hat{a}_\nu)^2]] = 0 \qquad (3.13-c)$$

$$\left[(\hat{a}_\iota{}^\dagger)^2, \left[(\hat{a}_\iota{}^\dagger)^2, (\hat{a}_\nu{}^\dagger)^2\right]\right] = \left[(\hat{a}_\nu{}^\dagger)^2, \left[(\hat{a}_\iota{}^\dagger)^2, (\hat{a}_\nu{}^\dagger)^2\right]\right] = 0 \qquad (3.13-d)$$

$$\iota, \nu = x, y, z$$
$$\iota \neq \nu$$

From equation (3.13) and after using (3.12), and the Baker-Campbell-Hausdorff relation one can easily show that the squeeze operator in 3D takes the more compact form

$$\hat{S}(\mathbf{s}) = \exp\left\{\frac{1}{2}\left[s_x(\hat{a}_x{}^\dagger)^2 + s_y(\hat{a}_y{}^\dagger)^2 + s_z(\hat{a}_z{}^\dagger)^2 - s_x{}^*(\hat{a}_x)^2 - s_y{}^*(\hat{a}_y)^2 - s_z{}^*(\hat{a}_z)^2\right]\right\} \qquad (3.14)$$

With the help of the definitions of vectors

$$\mathbf{s} = s_x \mathbf{i} + s_y \mathbf{j} + s_z \mathbf{k} \qquad (3.15-a)$$
$$\widehat{\mathbf{A}}^2 = \hat{a}_x{}^2 \mathbf{i} + \hat{a}_y{}^2 \mathbf{j} + \hat{a}_z{}^2 \mathbf{k} \qquad (3.15-b)$$
$$\widehat{\mathbf{A}}^{\dagger 2} = \hat{a}_x{}^{\dagger 2} \mathbf{i} + \hat{a}_y{}^{\dagger 2} \mathbf{j} + \hat{a}_z{}^{\dagger 2} \mathbf{k} \qquad (3.15-c)$$

we can find a rather compact form for squeezing operator as

$$\hat{S}(\mathbf{s}) = \exp\left[\frac{1}{2}\left(\mathbf{s} \cdot \widehat{\mathbf{A}}^{\dagger 2} - \mathbf{s}^* \cdot \widehat{\mathbf{A}}^2\right)\right] \qquad (3.16)$$

The following commutation relations clearly hold

$$[\hat{\iota}, \hat{\nu}] = [\hat{p}_\iota, \hat{p}_\nu] = [\hat{\partial}_\iota, \hat{\partial}_\nu] = [\hat{\iota}^2, \hat{\nu}^2] = \left[\hat{\partial}_\iota{}^2, \hat{\partial}_\nu{}^2\right] = [\hat{\iota}\hat{\partial}_\iota, \hat{\nu}\hat{\partial}_\nu] = 0 \qquad (3.17)$$

$$\iota, \nu = x, y, z$$
$$\iota \neq \nu$$

from which the alternate form of the squeeze operator is obtained

$$\hat{S}(\mathbf{s}) = \exp\left[-\mathbf{s}_1 \cdot \hat{\mathcal{R}}\hat{\partial} + i\frac{1}{2}\mathbf{s}_2 \cdot (\hat{\mathcal{R}}^2 + \hat{\partial}^2)\right] \qquad (3.18)$$

In the last equation we have used the short-hand notations

$$\mathbf{s}_1 = \text{Re}\{\mathbf{s}\} \qquad (3.19-a)$$
$$\mathbf{s}_2 = \text{Im}\{\mathbf{s}\} \qquad (3.19-b)$$
$$\hat{\mathcal{R}}\hat{\partial} = \hat{x}\hat{\partial}_x\mathbf{i} + \hat{y}\hat{\partial}_y\mathbf{j} + \hat{z}\hat{\partial}_z\mathbf{k} \qquad (3.19-c)$$
$$\hat{\mathcal{R}}^2 = \hat{x}^2\mathbf{i} + \hat{y}^2\mathbf{j} + \hat{z}^2\mathbf{k} \qquad (3.19-d)$$
$$\hat{\partial}^2 = \hat{\partial}_x^2\mathbf{i} + \hat{\partial}_y^2\mathbf{j} + \hat{\partial}_z^2\mathbf{k} \qquad (3.19-e)$$

## 4. Construction of squeezed states

For the generation of squeezed state we must apply the squeeze operator and then coherent operator on the ground state of harmonic oscillator. Here in this process, we use the notation of [15] employed in (3.6). This results in

$$|\mathbf{s}, \boldsymbol{\alpha}\rangle = \hat{D}(\boldsymbol{\alpha})\hat{S}(\mathbf{s})|0\rangle \qquad (3.20)$$

Notice that $\hat{S}(\mathbf{s})|0\rangle$ represents the squeezed vacuum. Following the previous definitions and after some algebra we reach at the position representation of the squeezed state as

$$\Psi_{sq}(\mathbf{r}) = \langle \mathbf{r}|\mathbf{s}, \boldsymbol{\alpha}\rangle =$$

$$\frac{1}{\pi^{\frac{3}{4}}C} \exp\left(-\frac{i}{2}\mathbf{r}_0 \cdot \mathbf{p}_0\right)\exp(i\mathbf{r}\cdot\mathbf{p}_0)\exp\left[-(x-x_0)^2\left(\frac{1}{2g_xC_x^2} - ih_x\right)\right]$$

$$\exp\left[-(y-y_0)^2\left(\frac{1}{2g_yC_y^2} - ih_y\right)\right]\exp\left[-(z-z_0)^2\left(\frac{1}{2g_zC_z^2} - ih_z\right)\right] \qquad (3.21)$$

Here, $\mathbf{r}_0 = x_0\mathbf{i} + y_0\mathbf{j} + z_0\mathbf{k}$, $\mathbf{p}_0 = p_{0_x}\mathbf{i} + p_{0_y}\mathbf{j} + p_{0_z}\mathbf{k}$, and $C = C_xC_yC_z$, where

$$C_\iota = \sqrt{g_\iota(1 + 2ih_\iota)} \qquad (3.22-a)$$

$$h_\iota = \frac{s_{2_\iota}\sinh(r_\iota)}{2r_\iota\exp(r_\iota)} \qquad (3.22-b)$$

$$\iota = x, y, z$$

$s_{1_\iota}$ and $s_{2_\iota}$ are elements of the vectors $\mathbf{s}_1$ and $\mathbf{s}_2$ defined in (3.19), $\alpha = (\mathbf{r}_0 + i\mathbf{p}_0)/\sqrt{2}$, and

$$r_\iota = \sqrt{(s_{1_\iota})^2 + (s_{2_\iota})^2} \qquad (3.23-a)$$

$$g_\iota = \cosh(t_\iota) + \frac{s_{1_\iota}}{r_\iota}\sinh(t_\iota) = \exp(r_\iota)\cos^2\left(\frac{\theta_\iota}{2}\right) + \exp(-r_\iota)\sin^2\left(\frac{\theta_\iota}{2}\right) \qquad (3.23-b)$$

$$\theta_\iota = \tan^{-1}\left(\frac{s_{2_\iota}}{s_{1_\iota}}\right) \qquad (3.23-c)$$

$$\iota = x, y, z$$

For the detailed derivation of the (3.21), please refer to the Appendix B.

*4.1. Further properties of squeezed states*

In this section we will consider two important properties of squeezed states; quadrature squeezing parameter and Mandel's $Q$ parameter. For 1D squeezed states these two are defined as scalars, while for the proposed 3D states we define the generalized quadrature squeezing and Mandel's $Q$ parameters in the vector form. In the following we start with Mandel's $Q$ parameter.

Mandel's $Q$ "as a measure of departure of the variance of the photon number n from the variance of a Poisson process" was first proposed and calculated by Mandel [22, 23].

$$Q \equiv \frac{\langle\Delta\hat{n}^2\rangle - \langle\hat{n}\rangle}{\langle\hat{n}\rangle} \quad ; \quad \langle\Delta\hat{n}^2\rangle = \langle\hat{n}^2\rangle - \langle\hat{n}\rangle^2 \qquad (3.24)$$

For an arbitrary state, $Q$ can be negative, zero, or positive, which respectively infers a super-Poissonian, Poissonian or sub-Poissonian statistics [24]. It should be added here that Mandel has shown that, one should expect the squeezed states to show sub-Poissonian photon statistics through normal detection schemes [23].

For our 3D squeezed states, we define a vectorial Mandel's $Q$ parameter, $\mathbf{Q} = (Q_x, Q_y, Q_z)$ where $Q_\iota$ is the Mandel's $Q$ parameter related to squeezing in the $\iota$ direction. Note that the proposed squeezed state here can also be represented as the multiplication of three squeezed sates:

$$\Psi_{sq}(\mathbf{r}) = \Psi_{sq}(x, y, z) = \Psi_{sq_x}(x)\,\Psi_{sq_y}(y)\,\Psi_{sq_z}(z) \qquad (3.25)$$

$$\Psi_{sq_\iota}(j) = \frac{1}{\pi^{\frac{1}{4}} C} \exp\left(-\frac{i}{2}\iota_0 p_{0_\iota}\right) \exp(i\,\iota p_{0_\iota}) \exp\left[-(\iota - \iota_0)^2 \left(\frac{1}{2 g_\iota C_\iota^2} - ih_\iota\right)\right] \;;\; \iota = x, y, z \qquad (3.26)$$

Now by using the results of [24] for 1D squeezed state, we can show that:

$$Q_\iota = \frac{|\alpha_\iota|^2 (e^{2r_\iota} \cos^2 \delta_\iota + e^{-2r_\iota} \sin^2 \delta_\iota) + 2 \sinh^2 r_\iota \cosh^2 r_\iota}{|\alpha_\iota|^2 + \sinh^2 r_\iota} - 1;\; \iota = x, y, z \qquad (3.27)$$

$\theta_\iota$ and $r_\iota$ are defined in (3.23) and

$$|\alpha_\iota|^2 = \frac{1}{2}(\iota_0^2 + p_{0_\iota}^2) \qquad (3.28-a)$$

$$\phi_\iota = \tan^{-1}\left(\frac{p_{0_\iota}}{\iota_0}\right) \qquad (3.28-b)$$

$$\delta_\iota = \theta_\iota - \frac{\phi_\iota}{2} \qquad (3.28-c)$$

$$\iota = x, y, z$$

In this case, Mandel's $Q$ parameter for squeezing in each direction ($Q_\iota \;;\; \iota = x, y, z$) can be negative, zero or positive which means the statistics of squeezed states in that particular direction is super-Poissonian, Poissonian or sub-Poissonian respectively. Surface plots of the Mandel's $Q$ parameter are illustrated in Figure 1, as a function of $\delta$ and $r$, while retaining $\alpha$ as a constant. As it can be seen, there is no dependence on the angle $\delta$ when $\alpha = 0$. For this special case, one can easily check from (3.27) that $Q_\iota = \cosh^2 r_\iota + \sinh^2 r_\iota$.

Similarly, we can also define vectorial quadrature operator:

$$\hat{\mathbf{R}}_1 = \hat{X}_1 \mathbf{i} + \hat{Y}_1 \mathbf{j} + \hat{Z}_1 \mathbf{k} = \frac{1}{2}(\hat{\mathbf{a}} + \hat{\mathbf{a}}^\dagger) \qquad (3.29-a)$$

$$\widehat{\mathbf{R}}_2 = \hat{X}_2\mathbf{i} + \hat{Y}_2\mathbf{j} + \hat{Z}_2\mathbf{k} = \frac{1}{2i}(\hat{\mathbf{a}} - \hat{\mathbf{a}}^\dagger) \qquad (3.29-b)$$

where $\hat{\mathbf{a}}$ and $\hat{\mathbf{a}}^\dagger$ are defined in (2.13). In 1D squeezed state variances of quadrature operators are measures of squeezing. In fact for a 1D squeezed state with quadrature $\hat{I}_1 = 1/2(\hat{a} + \hat{a}^\dagger)$ and $\hat{I}_1 = 1/2i(\hat{a} - \hat{a}^\dagger)$ operators squeezing exists if [25]:

$$\langle \Delta \hat{I}_1^{\ 2} \rangle < \frac{1}{4} \quad or \quad \langle \Delta \hat{I}_2^{\ 2} \rangle < \frac{1}{4} \qquad (3.30)$$

Again for our 3D squeezed state, we can calculate variances of elements of vectorial quadrature operators from the results of [24] for 1D squeezed state:

$$\langle \Delta\widehat{\mathbf{R}}_1^{\ 2} \rangle = \left( \langle \Delta\hat{X}_1^{\ 2} \rangle, \langle \Delta\hat{Y}_1^{\ 2} \rangle, \langle \Delta\hat{Z}_1^{\ 2} \rangle \right) \qquad (3.31-a)$$

$$\langle \Delta\widehat{\mathbf{R}}_2^{\ 2} \rangle = \left( \langle \Delta\hat{X}_2^{\ 2} \rangle, \langle \Delta\hat{Y}_2^{\ 2} \rangle, \langle \Delta\hat{Z}_2^{\ 2} \rangle \right) \qquad (3.31-b)$$

$$\langle \Delta\hat{I}_1^{\ 2} \rangle = \frac{1}{4}\left[ e^{2r_\iota} \cos^2\left(\frac{\phi_\iota}{2}\right) + e^{-2r_\iota} \sin^2\left(\frac{\phi_\iota}{2}\right) \right] \qquad (3.31-c)$$

$$\langle \Delta\hat{I}_2^{\ 2} \rangle = \frac{1}{4}\left[ e^{2r_\iota} \sin^2\left(\frac{\phi_\iota}{2}\right) + e^{-2r_\iota} \cos^2\left(\frac{\phi_\iota}{2}\right) \right] \qquad (3.31-d)$$

$$I = X, Y, Z$$
$$\iota = x, y, z$$

Thus for 3D squeezed state, in direction $\iota = x, y, z$ squeezing exists if:

$$\langle \Delta\hat{I}_1^{\ 2} \rangle < \frac{1}{4} \quad or \quad \langle \Delta\hat{I}_2^{\ 2} \rangle < \frac{1}{4} \quad ; \quad I = X, Y, Z \qquad (3.32)$$

Plots of squeeze parameters (3.31c) and (3.31d) versus $\phi$ and $r$ are shown in Figs. 2 and 3, respectively as surface and contour diagrams. As it can be seen, squeezed states happen over the domains in which (3.32) holds, and any of the squeeze parameters fall under $\frac{1}{4}$. Evidently, the transformation $r \to -r$ switches the subplots for $\langle \Delta\hat{I}_1^{\ 2} \rangle$ and $\langle \Delta\hat{I}_2^{\ 2} \rangle$, due to the algebraic forms of the expressions (3.31c) and (3.31d).

Furthermore, Figure 4 shows the domain of squeezed $\langle\Delta\hat{I}_{1,2}{}^2\rangle < \frac{1}{4}$ versus de-squeezed $\langle\Delta\hat{I}_{1,2}{}^2\rangle > \frac{1}{4}$ states, respectively, filled in with color contours, and left blank. The borders could be explicitly found by solving (3.31c) and (3.31d) for $\langle\Delta\hat{I}_{1,2}{}^2\rangle = \frac{1}{4}$. This gives after simplifications to the fairly compact expression

$$e^{\pm 2r_\iota} = \tan^2\left(\frac{\phi_\iota}{2}\right); \iota = x, y, z \qquad (3.32)$$

which defines the borders separating the squeezed and de-squeezed states.

## 5. Conclusions

In this paper, we presented new closed-form expressions for coherent states and squeeze operators of a generalized harmonic oscillator potential in three spatial dimensions. We defined proper creation and annihilation operators and succeeded in presenting simple expressions for the corresponding displacement and squeeze operators.

## Appendix A: Derivation of Wigner function of 3D harmonic oscillator

The position representation of $|m, n, l\rangle$ state 3D harmonic oscillator reads:

$$\Psi_{nml}(\mathbf{r}) = \langle\mathbf{r}|m, n, l\rangle = \frac{1}{\sqrt{2^{n+m+l}n!\,m!\,l!}}\left(\frac{\kappa^2}{\pi}\right)^{\frac{3}{4}}\exp\left(-\frac{1}{2}\kappa^2 r^2\right)H_n(\kappa x)H_m(\kappa y)H_l(\kappa z) \quad (A.1)$$

Placing the above in the definition of Wigner function in (2.5) gives:

$$W_{|m,n,l\rangle}(\mathbf{r}, \mathbf{p}) = \left(\frac{1}{2\pi\hbar}\right)^3 \frac{1}{2^{n+m+l}n!\,m!\,l!}\left(\frac{\kappa^2}{\pi}\right)^{\frac{3}{2}} \iiint_{-\infty}^{\infty} d^3\zeta \left\{\exp\left(-\frac{i}{\hbar}\mathbf{p}\cdot\boldsymbol{\zeta}\right)\exp\left(-\frac{1}{2}\kappa^2\left|\mathbf{r}-\tfrac{1}{2}\boldsymbol{\zeta}\right|^2\right)\exp\left(-\frac{1}{2}\kappa^2\left|\mathbf{r}+\tfrac{1}{2}\boldsymbol{\zeta}\right|^2\right)H_n\left[\kappa\left(x-\tfrac{1}{2}\zeta_x\right)\right]H_n\left[\kappa\left(x+\tfrac{1}{2}\zeta_x\right)\right]H_m\left[\kappa\left(y-\tfrac{1}{2}\zeta_y\right)\right]H_m\left[\kappa\left(y+\tfrac{1}{2}\zeta_y\right)\right]H_l\left[\kappa\left(z-\tfrac{1}{2}\zeta_z\right)\right]H_l\left[\kappa\left(z+\tfrac{1}{2}\zeta_z\right)\right]\right\}$$

$$= \left(\frac{1}{\pi\hbar}\right)^3 I_x \cdot I_y \cdot I_z \qquad (A.2)$$

$$I_x = \frac{1}{2^{n+1}n!}\sqrt{\frac{\kappa^2}{\pi}} \int_{-\infty}^{\infty} d\zeta_x \left\{ \exp\left(-\frac{i}{\hbar}p_x\zeta_x\right) \exp\left[-\frac{1}{2}\kappa^2\left(x-\frac{1}{2}\zeta_x\right)^2\right] \exp\left[-\frac{1}{2}\kappa^2\left(x+\frac{1}{2}\zeta_x\right)^2\right] H_n\left[\kappa\left(x-\frac{1}{2}\zeta_x\right)\right] H_n\left[\kappa\left(x+\frac{1}{2}\zeta_x\right)\right] \right\} \quad (A.3)$$

$$I_y = \frac{1}{2^{m+1}m!}\sqrt{\frac{\kappa^2}{\pi}} \int_{-\infty}^{\infty} d\zeta_y \left\{ \exp\left(-\frac{i}{\hbar}p_y\zeta_y\right) \exp\left[-\frac{1}{2}\kappa^2\left(y-\frac{1}{2}\zeta_y\right)^2\right] \exp\left[-\frac{1}{2}\kappa^2\left(y+\frac{1}{2}\zeta_y\right)^2\right] H_m\left[\kappa\left(y-\frac{1}{2}\zeta_y\right)\right] H_m\left[\kappa\left(y+\frac{1}{2}\zeta_y\right)\right] \right\} \quad (A.4)$$

$$I_z = \frac{1}{2^{l+1}l!}\sqrt{\frac{\kappa^2}{\pi}} \int_{-\infty}^{\infty} d\zeta_z \left\{ \exp\left(-\frac{i}{\hbar}p_z\zeta_z\right) \exp\left[-\frac{1}{2}\kappa^2\left(z-\frac{1}{2}\zeta_z\right)^2\right] \exp\left[-\frac{1}{2}\kappa^2\left(z+\frac{1}{2}\zeta_z\right)^2\right] H_l\left[\kappa\left(z-\frac{1}{2}\zeta_z\right)\right] H_l\left[\kappa\left(z+\frac{1}{2}\zeta_z\right)\right] \right\} \quad (A.5)$$

Consider for example the first integral $I_x$. By changing $\kappa\zeta_x \to \zeta_x$ we have:

$$I_x = \frac{\exp(-\kappa^2 x^2)}{\sqrt{\pi}\, 2^{n+1}n!} \int_{-\infty}^{\infty} d\zeta_x \left\{ \exp\left(-\frac{i}{\hbar\kappa}p_x\zeta_x - \frac{1}{4}\zeta_x^2\right) H_n\left(\kappa x - \frac{1}{2}\zeta_x\right) H_n\left(\kappa x + \frac{1}{2}\zeta_x\right) \right\} \quad (A.6)$$

and now by using the algebraic manipulation:

$$-\frac{1}{4}\zeta_x^2 - \frac{i}{\hbar\kappa}p_x\zeta_x = -\left(\frac{1}{2}\zeta_x\right)^2 - 2\frac{\zeta_x}{2}\left(\frac{ip_x}{\hbar\kappa}\right) - \left(\frac{ip_x}{\hbar\kappa}\right)^2 - \left(\frac{p_x}{\hbar\kappa}\right)^2 = -\left(\frac{\zeta_x}{2} + i\frac{p_x}{\hbar\kappa}\right)^2 - \left(\frac{p_x}{\hbar\kappa}\right)^2 \quad (A.7)$$

and change of variables $\left(\frac{\zeta_x}{2} + i\frac{p_x}{\hbar\kappa}\right) \to \xi_x$:

$$I_x = \frac{\exp\left[-(\kappa x)^2 - \left(\frac{p_x}{\hbar\kappa}\right)^2\right]}{2^n n!} \frac{1}{\sqrt{\pi}} \int_{-\infty}^{\infty} d\xi_x \left\{ \exp(-\xi_x^2) H_n\left(\kappa x + \xi_x - i\frac{p_x}{\hbar\kappa}\right) H_n\left(\kappa x - \xi_x + i\frac{p_x}{\hbar\kappa}\right) \right\} \quad (A.8)$$

from symmetry of Hermite polynomials we know that $H_n(-\xi) = (-1)^n H_n(\xi)$. So

$$I_x = (-1)^n \frac{\exp\left[-(\kappa x)^2 - \left(\frac{p_x}{\hbar \kappa}\right)^2\right]}{2^n n!} \frac{1}{\sqrt{\pi}}$$

$$\int_{-\infty}^{\infty} d\xi_x \left\{\exp(-\xi_x^2) H_n\left(\xi_x - i\frac{p_x}{\hbar \kappa} + \kappa x\right) H_n\left(\xi_x - i\frac{p_x}{\hbar \kappa} - \kappa x\right)\right\} \quad (A.9)$$

Also it is known that:

$$\frac{1}{2^n n!} \frac{1}{\sqrt{\pi}} \int_{-\infty}^{\infty} d\xi \{\exp(-\xi^2) H_n(\xi + \xi_1) H_n(\xi + \xi_2)\} = L_n(-2\xi_1 \xi_2) \quad (A.10)$$

where $L_n$ is the Laguerre polynomial of order $n$. Therefore

$$I_x = (-1)^n \exp\left[-(\kappa x)^2 - \left(\frac{p_x}{\hbar \kappa}\right)^2\right] L_n\left\{2\left[\left(\frac{p_x}{\hbar \kappa}\right)^2 + (\kappa x)^2\right]\right\} \quad (A.11)$$

Repeating the same procedure for $I_y$ and $I_z$ results in:

$$W_{|n,m,l\rangle}(\mathbf{r}, \mathbf{p}) = \frac{(-1)^{n+m+l}}{(\pi \hbar)^3} \exp\left[-\left(\frac{\mathbf{p}}{\hbar \kappa}\right)^2 - (\kappa \mathbf{r})^2\right]$$

$$L_n\left\{2\left[\left(\frac{p_x}{\hbar \kappa}\right)^2 + (\kappa x)^2\right]\right\} L_m\left\{2\left[\left(\frac{p_y}{\hbar \kappa}\right)^2 + (\kappa y)^2\right]\right\} L_l\left\{2\left[\left(\frac{p_z}{\hbar \kappa}\right)^2 + (\kappa z)^2\right]\right\} \quad (A.12)$$

**Appendix B: Derivation of Position representation of 3D squeezed state**

From (2.8) and by using the notation of [15] we can write the 3D displacement operator in this new form:

$$\hat{D}(\boldsymbol{\alpha}) = \hat{D}_x(\alpha_x) \hat{D}_y(\alpha_y) \hat{D}_z(\alpha_z)$$

$$\hat{D}_\iota(\alpha_\iota) = \exp\left(-\frac{i}{2} \iota_0 p_{0_\iota}\right) \exp(i p_{0_\iota} \hat{\iota}) \exp(-\iota_0 \hat{\partial}_\iota) \quad ; \quad \iota = x, y, z \quad (B.1)$$

where $\hat{\iota}$ and $\hat{\partial}_\iota$ are defined in (3.11) and $\iota_0$ and $p_{0_\iota}$ are define above the (3.23). Furthermore as it is shown in (3.10) 3D squeeze operator can also be shown to be:

$$\hat{S}(\mathbf{s}) = \hat{S}_x(s_x)\hat{S}_y(s_y)\hat{S}_z(s_z)$$

$$\hat{S}_\iota(s_\iota) = \exp\left[-s_{1_\iota}\hat{\iota}\hat{\partial}_\iota + i\frac{1}{2}s_{2_\iota}\left(\hat{\iota}^2 + \hat{\partial}_\iota^2\right)\right] \quad ; \quad \iota = x, y, z \qquad (B.2)$$

Our proposed squeezed state is constructed from ground state of a 3D harmonic oscillator (vacuum state) as in (3.20). So its position representation can be calculated from:

$$\Psi_{sq}(\mathbf{r}) = \langle \mathbf{r}|\mathbf{s}, \boldsymbol{\alpha} \rangle = \widehat{D}(\boldsymbol{\alpha})\hat{S}(\mathbf{s})\langle \mathbf{r}|0\rangle$$

$$= \widehat{D}(\boldsymbol{\alpha})\hat{S}(\mathbf{s}) \left(\frac{1}{\pi}\right)^{\frac{3}{4}} \exp\left[-\frac{1}{2}(x^2 + y^2 + z^2)\right] \qquad (B.3)$$

From the previously used commutation relation it is obvious that:

$$\Psi_{sq}(\mathbf{r}) = \left(\frac{1}{\pi}\right)^{\frac{3}{4}} \left[\widehat{D}_x(\alpha_x)\hat{S}_x(s_x)\exp\left(-\frac{1}{2}x^2\right)\right]\left[\widehat{D}_y(\alpha_y)\hat{S}_y(s_y)\exp\left(-\frac{1}{2}y^2\right)\right]$$

$$\left[\widehat{D}_z(\alpha_z)\hat{S}_z(s_z)\exp\left(-\frac{1}{2}z^2\right)\right] \qquad (B.4)$$

Using [15] gives:

$$\widehat{D}_\iota(\alpha_\iota)\hat{S}_\iota(s_\iota)\exp\left(-\frac{1}{2}\iota^2\right) = \frac{1}{C_\iota}\exp\left(-\frac{i}{2}\iota_0 p_{0_\iota}\right)\exp(i\,\iota p_{0_\iota})\exp\left[-(\iota - \iota_0)^2\left(\frac{1}{2\mathcal{G}_\iota C_\iota^2} - ih_\iota\right)\right] \qquad (B.5)$$

$$\iota = x, y, z$$

which directly results in the position representation of the squeezed state as:

$$\Psi_{sq}(\mathbf{r}) = \frac{1}{\pi^{\frac{3}{4}}C}\exp\left(-\frac{i}{2}\mathbf{r}_0 \cdot \mathbf{p}_0\right)\exp(i\mathbf{r} \cdot \mathbf{p}_0)\exp\left[-(x - x_0)^2\left(\frac{1}{2\mathcal{G}_x C_x^2} - ih_x\right)\right]$$

$$\exp\left[-(y - y_0)^2\left(\frac{1}{2\mathcal{G}_y C_y^2} - ih_y\right)\right]\exp\left[-(z - z_0)^2\left(\frac{1}{2\mathcal{G}_z C_z^2} - ih_z\right)\right] \qquad (B.6)$$

**Figure Captions**

Figure 1. Mandel's $Q$ parameter plotted versus $\delta$ and $r$ for various values of $\alpha$.

Figure 2. Squeeze parameters versus $\phi$ and $r$.

Figure 3. Contours of Squeeze parameters versus $\phi$ and $r$.

Figure 4. The domain of squeezed states (filled with contours) as separated from de-squeezed states (white).

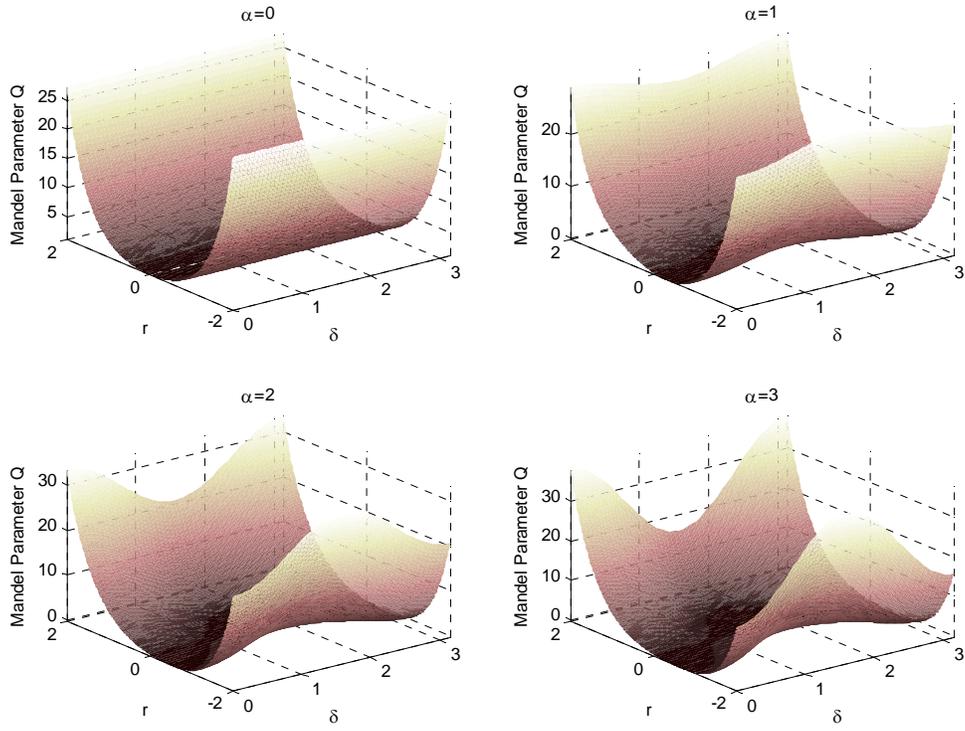

Figure 1

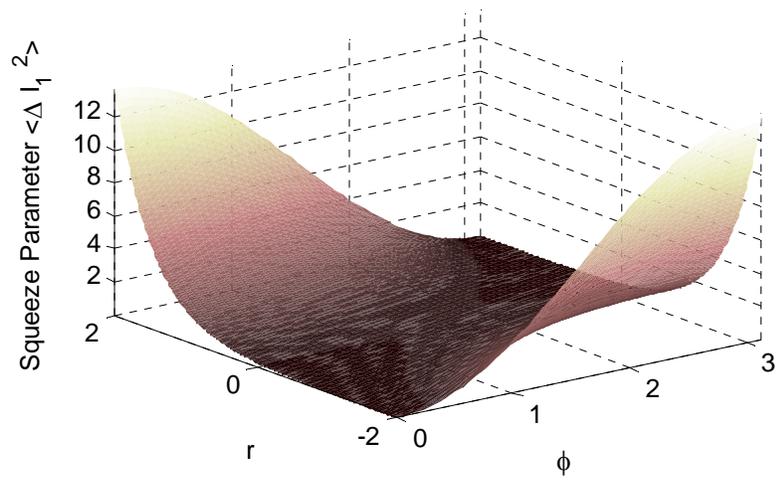

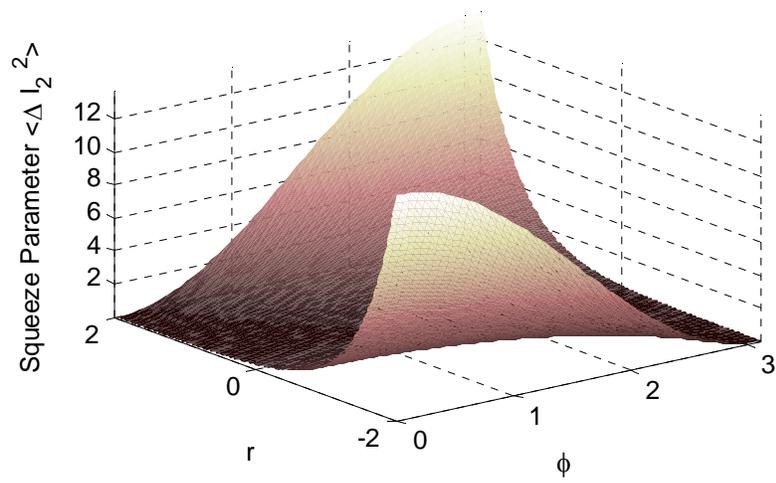

Figure 2

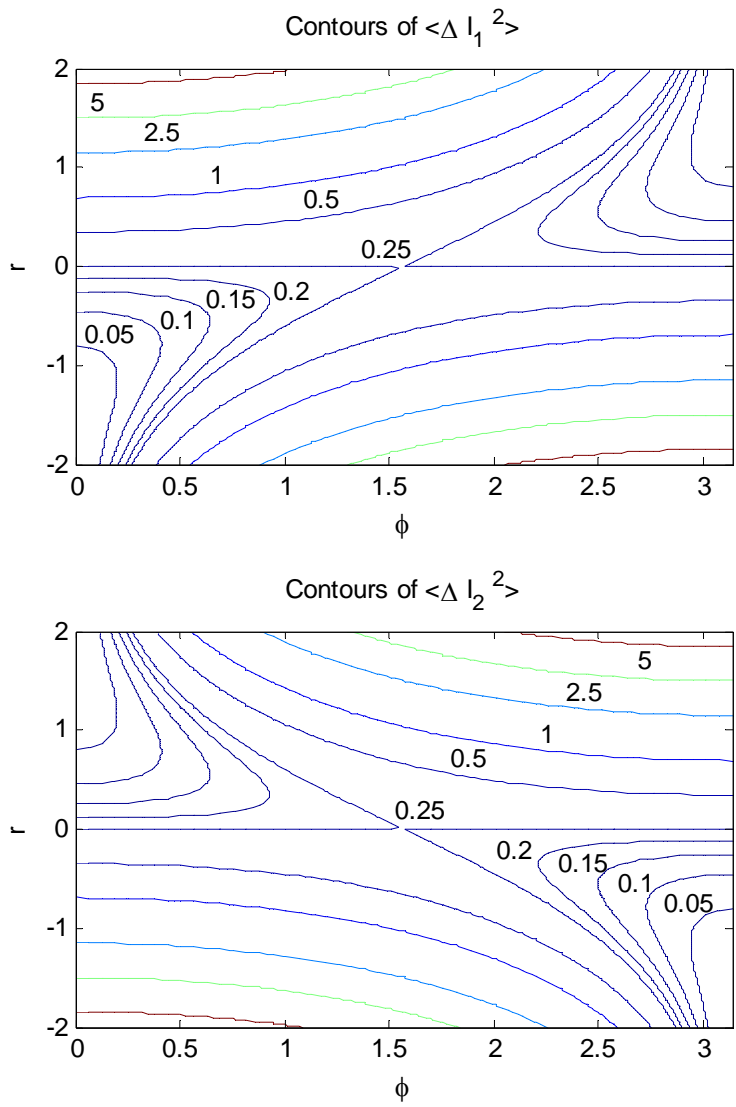

Figure 3

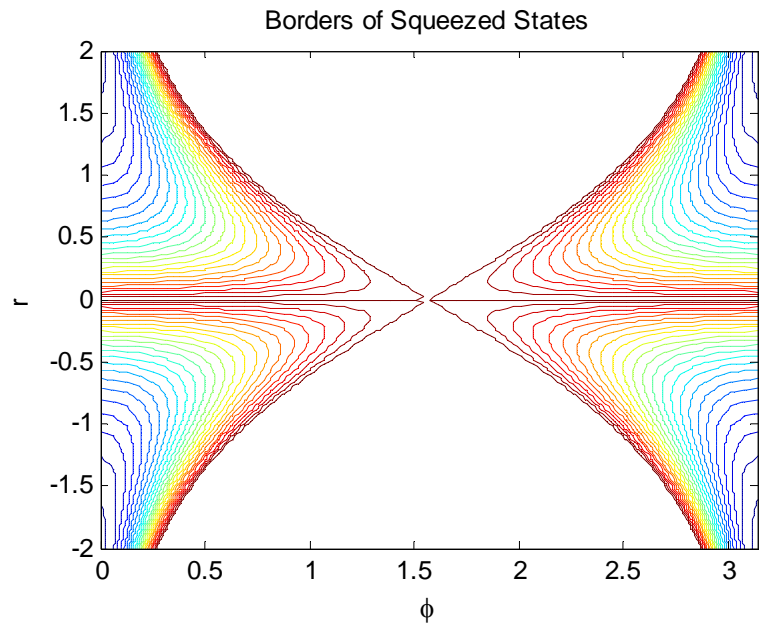

Figure 4